\documentclass[aps,prd,reprint,preprintnumbers,nofootinbib]{revtex4-1}

\usepackage[utf8]{inputenc}
\usepackage{graphicx}
\usepackage{bbold}
\usepackage{mathtools}

\usepackage[colorlinks,pdfusetitle]{hyperref}
\hypersetup{allcolors=[rgb]{0,0,1}}

\newcommand{\wh}[1]{\widehat{#1}}
\newcommand{\wt}[1]{\widetilde{#1}}
\newcommand{\h}{\mathcal H}
\newcommand{\diag}{\text{diag}}
\newcommand{\Dmsqee}{\Delta m^2_{ee}}
\newcommand{\sign}{\text{sign}}
\newcommand{\eps}{\epsilon}
\newcommand{\orcid}[1]{\href{https://orcid.org/#1}{#1}}
\newcommand{\e}[1]{\times10^{#1}}

\begin{document}

\title{Eigenvalues: the Rosetta Stone for Neutrino Oscillations in Matter}

\author{Peter B.~Denton}
\email{pdenton@bnl.gov}
\thanks{orcid \# \orcid{0000-0002-5209-872X}}
\affiliation{Physics Department, Brookhaven National Laboratory, Upton, NY 11973, USA}

\author{Stephen J.~Parke}
\email{parke@fnal.gov}
\thanks{orcid \# \orcid{0000-0003-2028-6782}}
\affiliation{Theoretical Physics Department, Fermi National Accelerator Laboratory, Batavia, IL 60510, USA}

\author{Xining Zhang}
\email{xining@uchicago.edu}
\thanks{orcid \# \orcid{0000-0001-8959-8405}}
\affiliation{Enrico Fermi Inst. and Dept. of Physics, University of Chicago, Chicago, IL 60637, USA}

\preprint{FERMILAB-PUB-19-326-T}

\date{May 1, 2020}

\begin{abstract}
We present a new method of exactly calculating neutrino oscillation probabilities in matter.
We leverage the ``eigenvector-eigenvalue identity'' to show that, given the eigenvalues, all mixing angles in matter follow surprisingly simply. The CP violating phase in matter can then be determined from the Toshev identity.
Then, to avoid the cumbersome expressions for the exact eigenvalues, we have applied previously derived perturbative, approximate eigenvalues to this scheme and discovered them to be even more precise than previously realized.
We also find that these eigenvalues converge at a rate of five orders of magnitude per perturbative order which is the square of the previously realized expectation.  Finally, we provide an updated speed versus accuracy plot for oscillation probabilities in matter, to include the methods of this paper.
\end{abstract}

\maketitle

\section{Introduction}
While the majority of the parameters in the three-neutrino oscillation picture have been measured, measurements of the remaining parameters will come by leveraging the matter effect in long-baseline experiments such as the currently running T2K and NOvA experiments, the now funded and under construction T2HK and DUNE experiments and the proposed T2HKK and ESSnuSB experiments, \cite{Ayres:2004js,Itow:2001ee,Acciarri:2016crz,Abe:2014oxa,Abe:2016ero,Baussan:2013zcy}.
In this context, only a full three-flavor picture including matter effects is adequate to probe the remaining parameters. 
Given the time and effort that is going into these experiments, it is paramount that we understand neutrino oscillations in matter as best as we can, both analytically and numerically, so as to maximize the oscillation physics output from these major experiments.

The matter effect is the fact that while neutrino propagation in vacuum occurs in the mass basis, in matter since the electron neutrino experiences an additional potential, they propagate in a new basis.
This effect was first identified in 1978 by Wolfenstein \cite{Wolfenstein:1977ue}.
Exact analytic solutions for neutrino oscillation probabilities in constant matter densities are difficult to fully enumerate; a solution using Lagrange's formula appeared in 1980 \cite{Barger:1980tf}\footnote{For more on Lagrange's formula in the context of neutrino oscillations in matter see ref.~\cite{Harrison:2002ee}.} while the full solution was first written down for three flavors in 1988 by Zaglauer and Schwarzer (ZS) \cite{Zaglauer:1988gz}.
The exact solution requires solving a cubic equation which, in the general case, has the unsightly and impenetrable $\cos(\frac13\cos^{-1}(\cdots))$ term present in the eigenvalues which are then in nearly every expression involving neutrino oscillations in matter\footnote{One interesting expression that does not contain the $\cos(\frac13\cos^{-1}(\cdots))$ term is the Jarlskog invariant in matter \cite{Denton:2019yiw}.}.
Given the eigenvalues, there are then several choices of how to map this onto the oscillation probabilities.
ZS mapped the eigenvalues onto the effective mixing angles and CP phase in matter; given the phase and the angles, it is then possible to write down the oscillation probabilities in matter using the vacuum expressions and the new phase, angles, and eigenvalues.
In 2002, Kimura, Takamura, and Yokomakura (KTY) presented a new mapping from the eigenvalues onto the oscillation probabilities by looking at the products of the lepton mixing matrix that actually appear in the probabilities \cite{Yokomakura:2000sv,Kimura:2002wd}, see also \cite{Aquino:1995,Aquino:1995kb,Fogli:2001wi,Galais:2011jh}.
Another formulation of the exact result in the context of the time evolution operator is ref.~\cite{Ohlsson:1999um}.
Along with these exact expressions, numerous approximate expressions have appeared in the literature in various attempts to avoid the $\cos(\frac13\cos^{-1}(\cdots))$ term, for a recent review see ref.~\cite{Parke:2019vbs}.

In this article we use the eigenvector-eigenvalue identity, that has been recently and extensively surveyed in \cite{Denton:2019pka,ttblog}, to write the exact expressions for the mixing angles in matter in terms of the eigenvalues of the Hamiltonian and its principal minors.
The benefit of this approach is two-fold.
First, it makes the expressions for the mixing angles in matter, clearer, symmetric, and very simple.
Since our approach for the oscillation probabilities in matter is based on the form of the vacuum expressions, the intuition that exists for the vacuum still applies in matter.
Second, it allows for a simple replacement of the complicated exact eigenvalues with far simpler approximate eigenvalues in a straightforward fashion.
We find that since this approximate approach only relies on approximate expressions for the eigenvalues, it is more accurate than previous methods, including Denton, Minakata and Parke (DMP) \cite{Denton:2016wmg}, with a comparable level of simplicity.
We also explore the convergence rate of the eigenvalues in DMP and find that since all odd-order corrections vanish, they converge much faster than expected, at a rate of $\sim10^{-5}$.

\section{An Eigenvalue Based Exact Solution}
The technique of ZS is to determine expressions for the mixing angles and CP-violating phase in matter ($\wh\theta_{23}$, $\wh\theta_{13}$, $\wh\theta_{12}$, and $\wh\delta$) as a function of the eigenvalues and other expressions, while KTY derives the general expression for the product of elements of the lepton mixing matrix, $U_{\alpha i}U_{\beta j}^*$.
In this section, we describe a technique of using both approaches.

First, we note that, given the eigenvalues, the mixing angles can be determined from various $|U_{\alpha i}|^2$ terms.
This employs a simpler version of the main result of KTY.
Then, to address the CP-violation part of the oscillation probabilities, we use the Toshev identity \cite{Toshev:1991ku}.

\subsection{Mixing Angles in Matter}
The neutrino oscillation Hamiltonian in matter in the flavor basis is
\begin{align}
&H= \nonumber \\[1mm]
&\frac1{2E}\left[U_{\rm PMNS}
\begin{pmatrix}
0\\&\Delta m^2_{21}\\&&\Delta m^2_{31}
\end{pmatrix}
U_{\rm PMNS}^\dagger+
\begin{pmatrix}
a&&\\&0\\&&0
\end{pmatrix}\right]\,,
\label{eq:H}
\end{align}
where we have subtracted out an overall $\frac{m_1^2}{2E}\mathbb1$, $a\equiv2\sqrt2G_Fn_eE$ is the Wolfenstein matter potential \cite{Wolfenstein:1977ue}, and the PMNS lepton mixing matrix \cite{Pontecorvo:1967fh,Maki:1962mu} is parameterized,
\begin{align}
&U_{\rm PMNS}= \nonumber \\[1mm]
&\begin{pmatrix}
1\\&c_{23}&s_{23}e^{i\delta}\\&-s_{23}e^{-i\delta}&c_{23}
\end{pmatrix}
\begin{pmatrix}
c_{13}&&s_{13}\\&1\\-s_{13}&&c_{13}
\end{pmatrix}
\begin{pmatrix}
c_{12}&s_{12}\\
-s_{12}&c_{12}\\
&&1
\end{pmatrix}\,, 
\end{align}
where $s_{ij}=\sin\theta_{ij}$, $c_{ij}=\cos\theta_{ij}$, and we have shifted the CP-violating phase $\delta$ from its usual position on $s_{13}$ to $s_{23}$ which does not affect any observable.
For our numerical studies we use $\Delta m^2_{21}=7.55\e{-5}$ eV$^2$, $\Delta m^2_{31}=2.5\e{-3}$ eV$^2$, $s_{12}^2=0.32$, $s_{13}^2=0.0216$, $s_{23}^2=0.547$, and $\delta=1.32\pi$ from \cite{deSalas:2017kay}.

Using the eigenvector-eigenvalue identity \cite{Denton:2019pka}, the squares of the elements of the lepton mixing matrix in matter are simple functions of the eigenvalues of the neutrino oscillation Hamiltonian in matter, $\lambda_i/2E$ for $i\in\{1,2,3\}$, and new submatrix eigenvalues, $\xi_\alpha/2E$ and $\chi_\alpha/2E$ for $\alpha\in\{e,\mu,\tau\}$.
In general, the square of the elements of the mixing matrix are parameterization independent,
\begin{equation}
|\wh U_{\alpha i}|^2=\frac{(\lambda_i-\xi_\alpha)(\lambda_i-\chi_\alpha)}{(\lambda_i-\lambda_j)(\lambda_i-\lambda_k)}\,,
\label{eq:Uaisq}
\end{equation}
where $i$, $j$, and $k$ are all different, and the $\lambda_i$ are the exact eigenvalues, see appendix \ref{sec:exact eigenvalues}.
This result, eq.~\ref{eq:Uaisq}, can also be directly obtained from KTY as shown in appendix \ref{sec:from kty}.
This equation is valid for every element of the mixing matrix, even the $\mu$ and $\tau$ rows, which are relatively complicated in the standard parameterization.

Eq.~\ref{eq:Uaisq} is one of the primary results of our paper.
Given the eigenvalues of the Hamiltonian and the eigenvalues of the submatrix Hamiltonian, it is possible to write down all nine elements of the mixing matrix in matter, squared.
This result is also quite simple and easy to memorize which is contrasted with the complicated forms from previous solutions \cite{Zaglauer:1988gz,Kimura:2002wd,Xing:2019owb}.

\begin{figure*}
\includegraphics[width=0.325\textwidth]{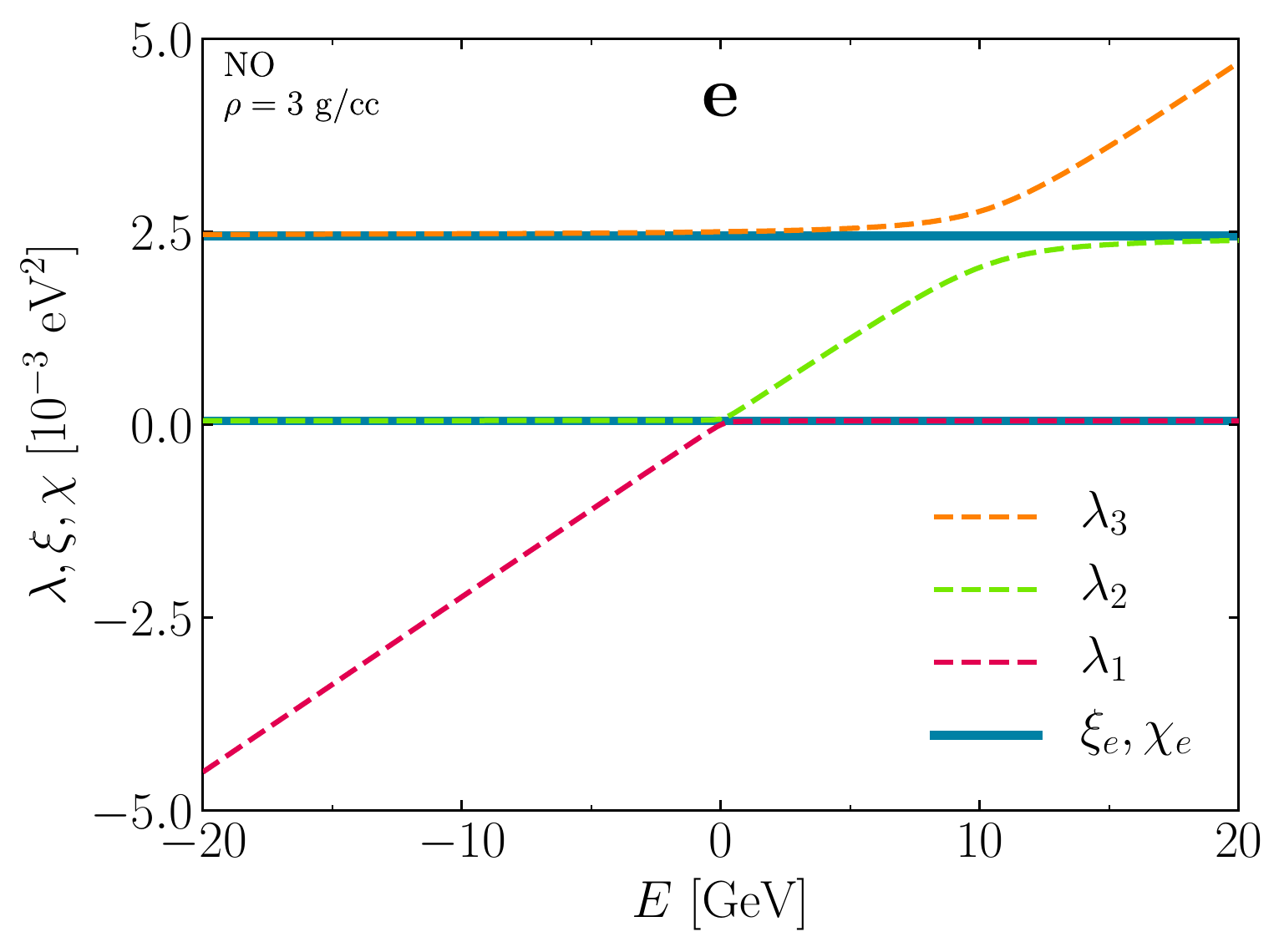}
\includegraphics[width=0.325\textwidth]{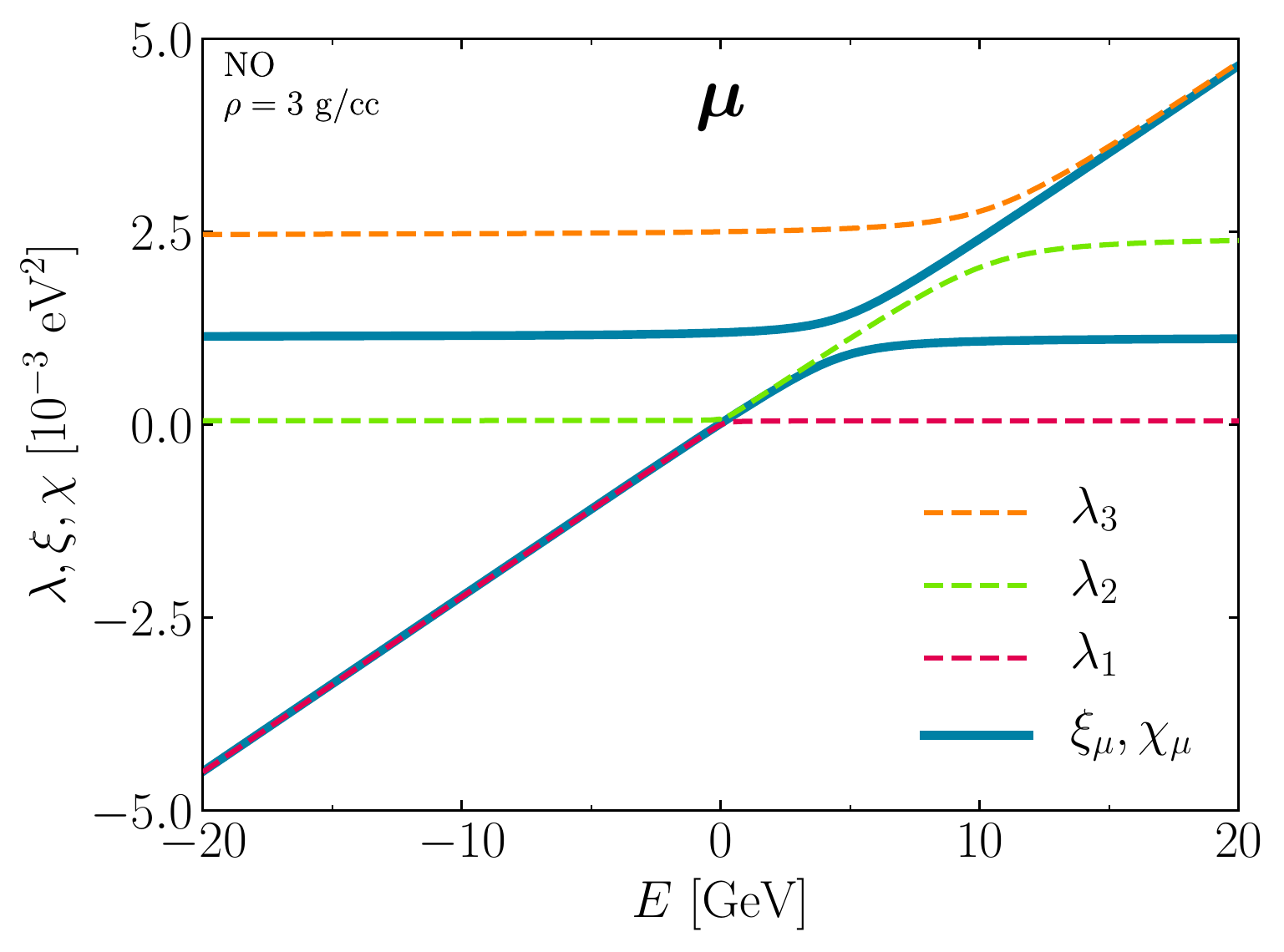}
\includegraphics[width=0.325\textwidth]{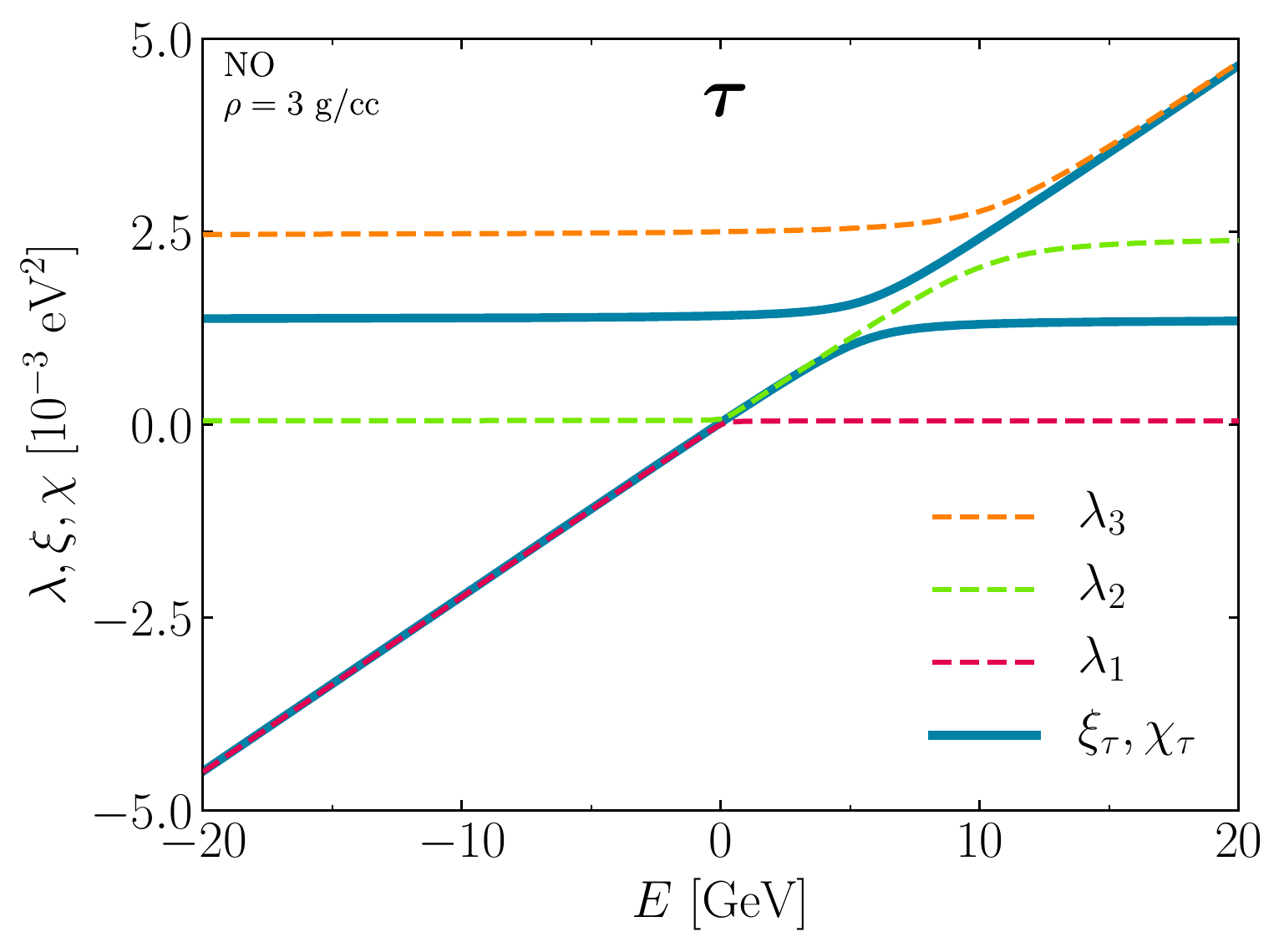}
\caption{
The two submatrix eigenvalues, $\xi_\alpha$ and $\chi_\alpha$,  as a function of neutrino energy, are shown in the solid blue curves with $\alpha=e,\mu,\tau$ in the left, center, and right figures respectively.
For comparison, the full matrix eigenvalues $\lambda_i$ are shown in dashed red, green, and orange in each panel.
When a submatrix eigenvalue (solid) overlaps one of the full matrix eigenvalues (dashed) the corresponding $|U_{\alpha i}|^2 \rightarrow 0$, as seen from the numerator of eq.~\ref{eq:Uaisq}.
Note the Cauchy interlacing condition is satisfied, $\lambda_1 \leq \xi_\alpha \leq \lambda_2 \leq \chi_\alpha \leq \lambda_3$, for each $\alpha=(e,\mu,\tau)$ and all $E$, using the convention  $\xi_\alpha < \chi_\alpha$.
See appendix \ref{sec:submatrix asymptotics} for further discussion.}
\label{fig:submatrix eigenvalues}
\end{figure*}

The submatrix eigenvalues $\xi_\alpha/2E$ and $\chi_\alpha/2E$ are the eigenvalues of the $2\times2$ submatrix of the Hamiltonian,
\begin{equation}
H_\alpha\equiv
\begin{pmatrix}
H_{\beta\beta}&H_{\beta\gamma}\\
H_{\gamma\beta}&H_{\gamma\gamma}
\end{pmatrix}\,,
\end{equation}
for $\alpha$, $\beta$, and $\gamma$ all different.
Explicit expressions for the Hamiltonian are given in appendix \ref{sec:hamiltonian} and the eigenvalues of the submatrices, which require only the solution to a quadratic, are plotted in fig.~\ref{fig:submatrix eigenvalues}.
We note that while solving a quadratic is necessary to evaluate the submatrix eigenvalues, since only the sum and the product of the eigenvalues (that is, the trace and the determinant of the submatrix Hamiltonian) appear in eq.~\ref{eq:Uaisq} whose numerator can be rewritten as $\lambda_i^2-\lambda_i(\xi_\alpha+\chi_\alpha)+\xi_\alpha\chi_\alpha$, the submatrix eigenvalues do not have to be explicitly calculated.
The expressions for the sums and products of the eigenvalues are given in appendix \ref{sec:hamiltonian}.

Given the standard parameterization of the lepton mixing matrix, this allows us to write all three mixing angles in matter as simple expressions of the eigenvalues,
\begin{align}
s_{\wh{12}}^2c_{\wh{13}}^2=|\wh U_{e2}|^2&=-\frac{(\lambda_2-\xi_e)(\lambda_2-\chi_e)}{(\lambda_3-\lambda_2)(\lambda_2-\lambda_1)}\,,\label{eq:s12mc13m}\\
s_{\wh{13}}^2=|\wh U_{e3}|^2&=\frac{(\lambda_3-\xi_e)(\lambda_3-\chi_e)}{(\lambda_3-\lambda_1)(\lambda_3-\lambda_2)}\,,\label{eq:s13m}\\
s_{\wh{23}}^2c_{\wh{13}}^2=|\wh U_{\mu3}|^2&=\frac{(\lambda_3-\xi_\mu)(\lambda_3-\chi_\mu)}{(\lambda_3-\lambda_1)(\lambda_3-\lambda_2)}\,,\label{eq:s23mc13m}
\end{align}
where the hat indicates that it is the mixing angle in matter.
While similar versions of eqs.~\ref{eq:s12mc13m} and \ref{eq:s13m} have previously appeared in the literature \cite{Zaglauer:1988gz}, eq.~\ref{eq:s23mc13m} is original to this manuscript.
The general form of eq.~\ref{eq:Uaisq} allows us to write any term in the lepton mixing matrix, and thus any mixing angle with considerable ease.
In addition, as we will show in the next section, this also allows us to calculate the CP violating phase quite easily.

In appendix \ref{sec:rotated basis} we show how to use this method in the vacuum $(\theta_{23}, \delta)$-rotated flavor basis. Further extensions of eq.~\ref{eq:Uaisq} to an arbitrary number of neutrinos is also discussed in appendix \ref{sec:extension arbitrary}.

\subsection{CP-Violating Phase in Matter}
In order to determine the CP-violating phase in matter, we note that $\cos\wh\delta$ can be determined, given the other mixing angles in matter, from $|U_{\mu1}|^2$ (or $|U_{\mu2}|^2$, $|U_{\tau1}|^2$, or $|U_{\tau2}|^2$) from eq.~\ref{eq:Uaisq}.
The sign of $\wh\delta$ needs to be separately determined.
We note that the sign of $\wh\delta$ must be the same as the sign of $\delta$.
To see this, we employ the Naumov-Harrison-Scott (NHS) identity \cite{Naumov:1991ju,Harrison:1999df},
\begin{equation}
\wh J=\frac{\Delta m^2_{21}\Delta m^2_{31}\Delta m^2_{32}}{\Delta\wh{m^2}_{21}\Delta\wh{m^2}_{31}\Delta\wh{m^2}_{32}}J\,,
\label{eq:NHS}
\end{equation}
where $J=\Im[U_{e1}U_{\mu2}U_{e2}^*U_{\mu1}^*]=s_{23}c_{23}s_{13}c_{13}^2s_{12}c_{12}\sin\delta$ is the Jarlskog invariant \cite{Jarlskog:1985ht}.
We note that the numerator and denominator in eq.~\ref{eq:NHS} always have the same sign, so $\sin\wh\delta$ has the same sign as $\sin\delta$.
That is, the eigenvalues in matter never cross.

In practice, it is simpler to determine $\sin\wh\delta$ from the Toshev identity \cite{Toshev:1991ku},
\begin{equation}
\sin\wh\delta=\frac{\sin2\theta_{23}}{\sin2\wh\theta_{23}}\sin\delta\,,
\label{eq:toshev}
\end{equation}
and use $\wh\theta_{23}$ determined in eq.~\ref{eq:s23mc13m}.
An alternative method for determining the CP violating phase is given in appendix \ref{sec:rotated basis}.

\subsection{Oscillation Probabilities in Matter}
Finally, these can all be combined into any oscillation probability.
For the primary appearance channel at NOVA, T2K, DUNE, T2HK(K), ESSnuSB \cite{Ayres:2004js,Itow:2001ee,Acciarri:2016crz,Abe:2014oxa,Abe:2016ero,Baussan:2013zcy}, or any other long-baseline neutrino experiment, the oscillation probability can be written in the following compact form,
\begin{equation}
P(\nu_\mu\to\nu_e)=\left|\mathcal A_{31}+e^{\pm i\Delta_{32}}\mathcal A_{21}\right|^2\,,
\end{equation}
where the upper (lower) sign is for neutrinos (anti-neutrinos) and
\begin{align}
\mathcal A_{31}&=2s_{\wh{13}}c_{\wh{13}}s_{\wh{23}}\sin\wh\Delta_{31}\,,\label{eq:A31}\\
\mathcal A_{21}&=2s_{\wh{12}}c_{\wh{13}}(c_{\wh{12}}c_{\wh{23}}e^{-i\wh\delta}-s_{\wh{12}}s_{\wh{13}}s_{\wh{23}})\sin\wh\Delta_{21}\,,\label{eq:A21}\\
\wh\Delta_{ij}&=\frac{(\lambda_i-\lambda_j)L}{4E}\,.\label{eq:Dij}
\end{align}
The above determination of the mixing angles and CP violating phase in matter allow for the simple determination of the oscillation probability in matter for the $\nu_\mu\to\nu_e$ appearance channel, or any other channel via the vacuum oscillation probabilities.
Therefore, any physics intuition already obtained for vacuum oscillation probabilities is easily transferred to oscillation probabilities in matter.
 
In addition to all the appearance channels this approach also works in a straightforward fashion for the disappearance channels as well.
For disappearance the oscillation probabilities in matter can be written as,
\begin{equation}
P(\nu_\alpha\to\nu_\alpha)=1-4\sum_{i<j}|\wh U_{\alpha i}|^2|\wh U_{\alpha j}|^2\sin^2\wh\Delta_{ij}\,.
\end{equation}
Thus the coefficients, $|\wh U_{\alpha i}|^2|\wh U_{\alpha j}|^2$, can be read off as simple functions of the eigenvalues and the submatrix eigenvalues, eq.~\ref{eq:Uaisq}, without any need to even convert to the mixing angles in matter. 
 
\section{Approximate Eigenvalues}
While the form of the mixing angles in matter presented above is exact, it still relies on the complicated expression of the eigenvalues.
It has been previously shown, however, that the eigenvalues can be extremely well approximated via a mechanism of changing bases as demonstrated by Denton, Minakata, and Parke (DMP) \cite{Denton:2016wmg}, see also refs.~\cite{Agarwalla:2013tza,Minakata:2015gra,Denton:2018fex}.
While expressions for the differences of eigenvalues in DMP are quite compact \cite{Denton:2018hal}, the expressions in eqs.~\ref{eq:s12mc13m}-\ref{eq:s23mc13m} require the individual eigenvalues so we list those here as well.
Beyond the zeroth order expressions, it is possible to derive higher order terms through perturbation theory \cite{Denton:2016wmg} or through further rotations \cite{Denton:2018fex}.
This approach leads to a smallness parameter that is no larger that $c_{12}s_{12}\frac{\Delta m^2_{21}}{\Delta m^2_{31}}\sim1.5\%$ and is zero in vacuum confirming that the exact solution is restored at zeroth order in vacuum, see eq.~\ref{eq:epsp} below.

\begin{figure}
\centering
\includegraphics[width=3.4in]{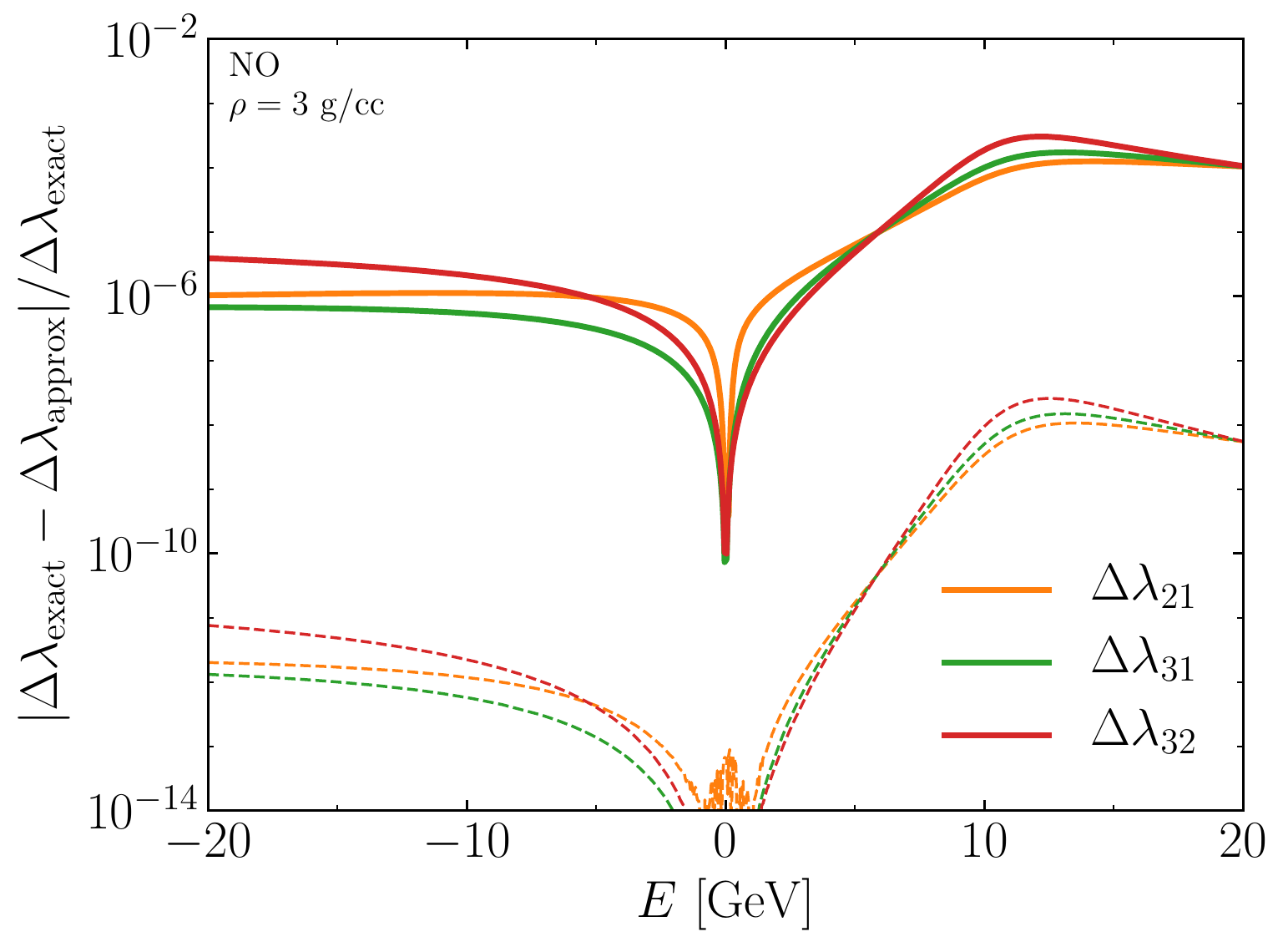}
\caption{The fractional precision of the zeroth order and the second order DMP eigenvalues are shown in solid and dashed curves respectively.
We plot the different in eigenvalues so that we are insensitive to an overall shift in the eigenvalues.
Continuing to higher order in the eigenvalues continues to increase the precision by comparable levels since all odd order corrections to the eigenvalues in DMP are zero (see appendix \ref{sec:high order eigenvalues}).}
\label{fig:eigenvalue precision}
\end{figure}

\subsection{Zeroth Order Eigenvalues}
The zeroth-order eigenvalues are extremely precise with a fractional error in the difference of the eigenvalues of $<10^{-5}$ at DUNE.
We define eigenvalues of two intermediate steps.
First, the eigenvalues of the un-rotated Hamiltonian, after a constant $U_{23}(\theta_{23},\delta)$ rotation\footnote{A term 
$(s^2_{12}\Delta m^2_{21}/2E)\mathbb1$ could be subtract from the Hamiltonian, eq.~\ref{eq:H}, to simplify the following expressions.},
\begin{align}
\lambda_a&=a+s_{13}^2\Dmsqee+s_{12}^2\Delta m^2_{21}\,,\\
\lambda_b&=c_{12}^2\Delta m^2_{21}\,,\\
\lambda_c&=c_{13}^2\Dmsqee+s_{12}^2\Delta m^2_{21}\,.
\end{align}
Next, after an $O_{13}$ rotation, we have
\begin{align}
\lambda_\pm&=\frac12\left[\lambda_a+\lambda_c \right. \notag \\ &  \quad \left. \pm\sign(\Dmsqee)\sqrt{(\lambda_a-\lambda_c)^2+(2s_{13}c_{13}\Dmsqee)^2}\right]\,,  \nonumber \\
\lambda_0&=\lambda_b\, ,
\label{eq:lambdapm dmp}
\end{align}
and
\begin{equation}
\sin^2\phi=\frac{\lambda_+-\lambda_c}{\lambda_+-\lambda_-}\,.
\label{eq:phi}
\end{equation}
Finally, after an $O_{12}$ rotation, the eigenvalues through zeroth order are
\begin{align}
\wt\lambda_{1,2}&=\frac12\left[\lambda_0+\lambda_-
 \right.  \label{eq:lambda12 dmp} \\ & \quad \left.\mp\sqrt{(\lambda_0-\lambda_-)^2+(2\cos(\phi-\theta_{13})s_{12}c_{12}\Delta m^2_{21})^2}\right]\,, \nonumber \\
\wt\lambda_3&=\lambda_+\,,\label{eq:lambda3 dmp}
\end{align}
and
\begin{equation}
\sin^2\psi=\frac{\wt\lambda_2-\lambda_0}{\wt\lambda_2-\wt\lambda_1}\,.
\label{eq:psi}
\end{equation}
Here $\wt x$ represents an approximate expressions for the quantity $x$ in matter. At this order, $\theta_{23}$ and $\delta$, are unchanged from their vacuum values.
Eqs.~\ref{eq:lambdapm dmp} to \ref{eq:psi} define the zeroth order approximation.

We note that $\phi$ and $\psi$ are an excellent approximation for $\wh\theta_{13}$ and  $\wh\theta_{12}$ respectively, \cite{Denton:2016wmg}, see ref.~\cite{Denton:2018fex} for the explicit higher order correction terms.
These  are  effective two-flavor approximations to $\wh\theta_{12}$ and  $\wh\theta_{13}$ while eqs.~\ref{eq:s12mc13m} and \ref{eq:s13m} are the full three-flavor exact expressions.
We further discuss the similarity in these expressions in sec.~\ref{sec:extension arbitrary}.

\subsection{Second Order Eigenvalues}
\label{sec:second order eigenvalues}
After performing the rotations that lead to the eigenvalues in eqs.~\ref{eq:lambda12 dmp} and \ref{eq:lambda3 dmp}, the smallness parameter is
\begin{equation}
\eps^\prime \,\equiv\sin(\phi-\theta_{13})s_{12}c_{12}\Delta m^2_{21}/\Dmsqee<1.5\%
\label{eq:epsp}
\end{equation}
 and is zero in vacuum since $\phi=\theta_{13}$ in vacuum.
Because of the nature of the DMP approximation, the zeroth order eigenvalues in eqs.~\ref{eq:lambda12 dmp} and \ref{eq:lambda3 dmp} already contain the first order in $\eps'$ corrections.
That is, the first order corrections are just the diagonal elements in the perturbing Hamiltonian which are all zero by construction.
The second order corrections are simply,
\begin{align}
\wt\lambda_1^{(2)}&=-(\eps'\Dmsqee)^2\frac{s_\psi^2}{\wt\lambda_3-\wt\lambda_1}\,,\label{eq:l12}\\
\wt\lambda_2^{(2)}&=-(\eps'\Dmsqee)^2\frac{c_\psi^2}{\wt\lambda_3-\wt\lambda_2}\,,\label{eq:l22}\\
\wt\lambda_3^{(2)}&=-\wt\lambda_1^{(2)}-\wt\lambda_2^{(2)}\,.\label{eq:l32}
\end{align}
It is useful to note that $\wt\lambda_1$ and $\wt\lambda_2$ are related by the 1-2 interchange symmetry \cite{Denton:2016wmg}.
The 1-2 interchange symmetry says that all oscillation observables in matter are independent of the following transformations,
\begin{equation}
\wt\lambda_1\leftrightarrow\wt\lambda_2\,,\quad
c_\psi^2\leftrightarrow s_\psi^2\,,\qquad\text{and}\quad
c_\psi s_\psi\to-c_\psi s_\psi\,.\label{eq:12 interchange symmetry}
\end{equation}
It is clear that $\wt\lambda^{(2)}_3$ is invariant under this interchange, and $\wt\lambda^{(2)}_2$ follows directly from $\wt\lambda^{(2)}_1$ and the interchange.

\begin{figure*}[t]
\centering
\includegraphics[width=0.45\textwidth]{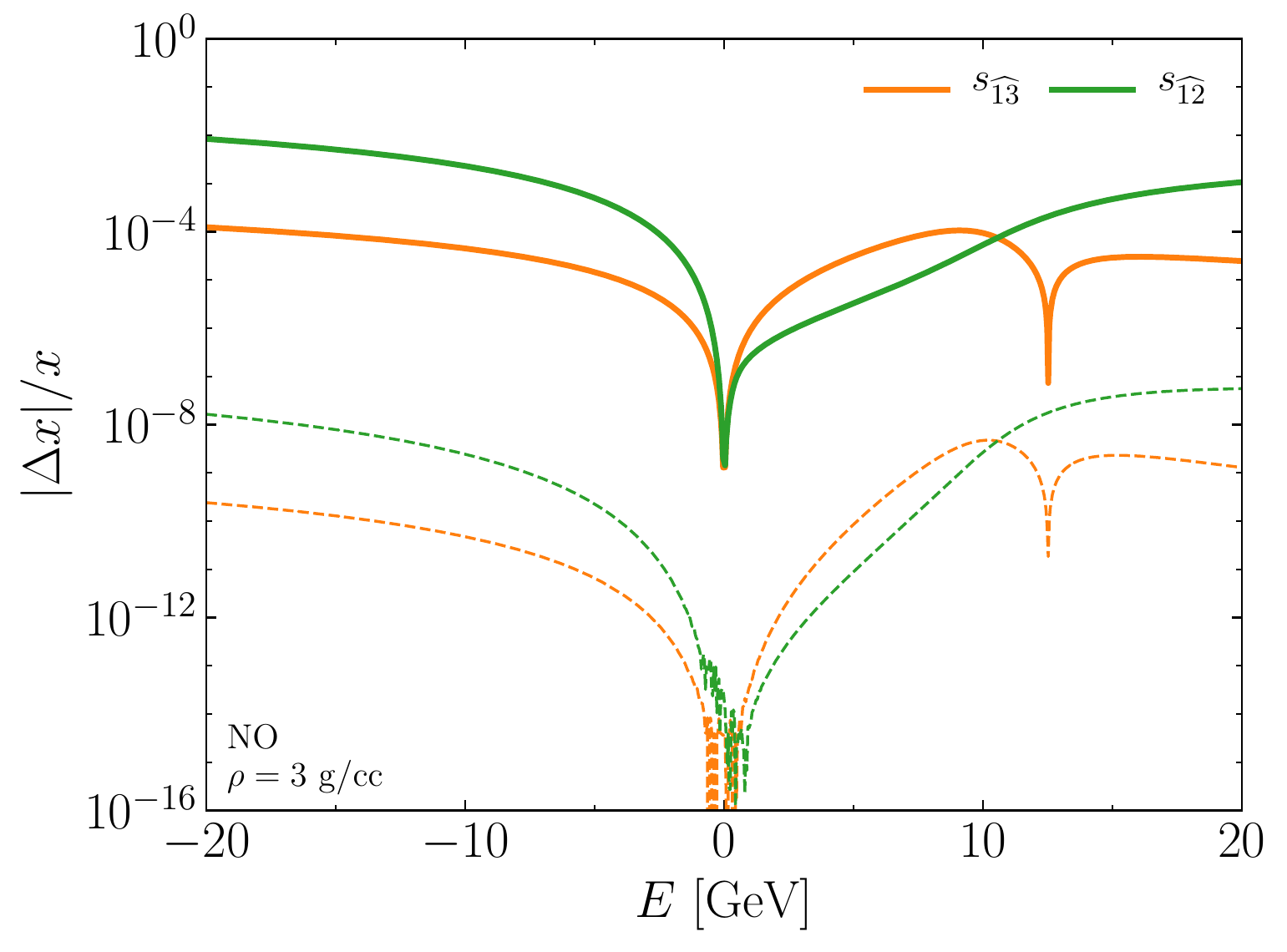}
\includegraphics[width=0.45\textwidth]{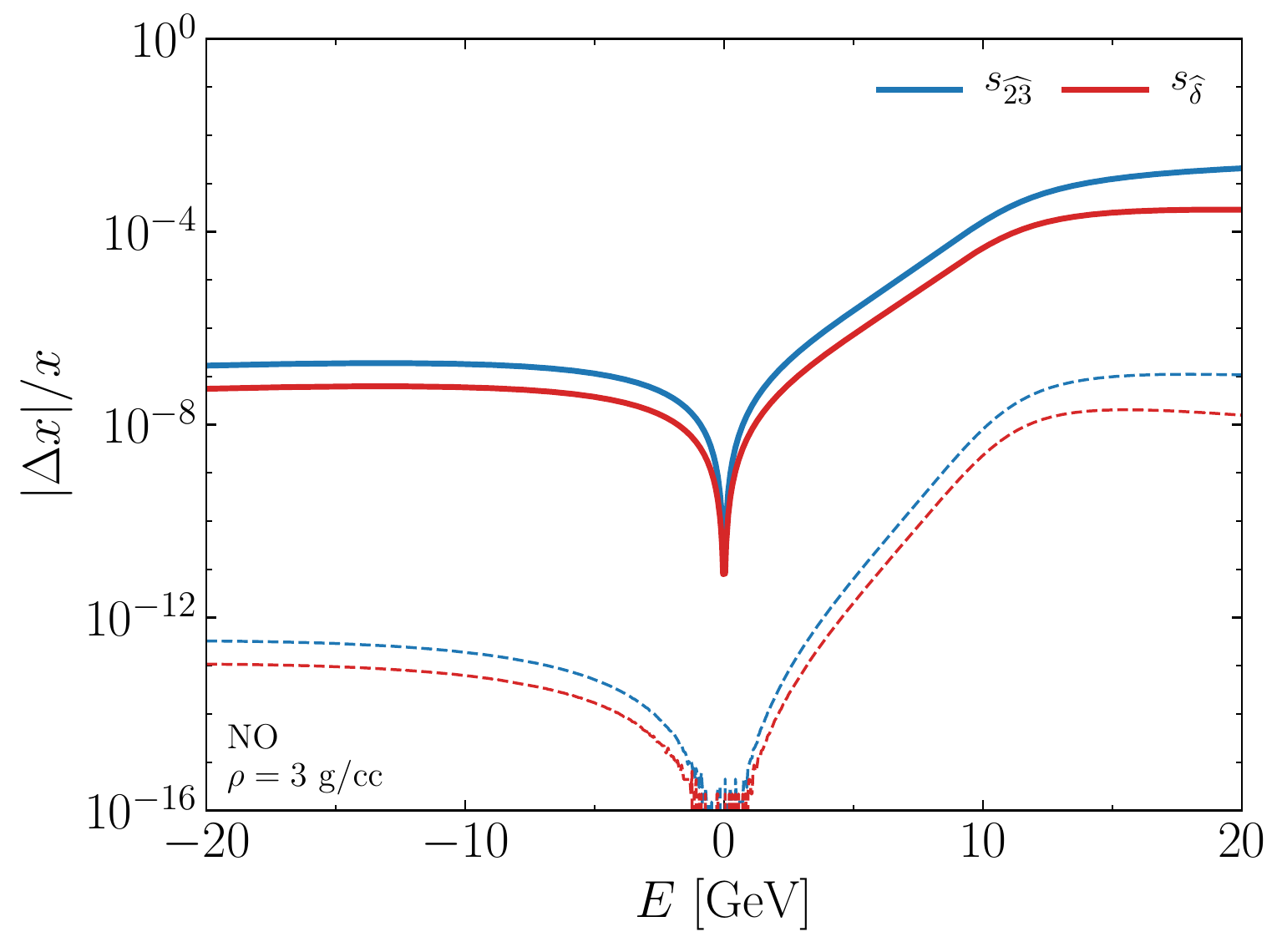}
\caption{The fractional precision of sine of the mixing angles and the CP violating phase in matter in eqs.~\ref{eq:s12mc13m}-\ref{eq:toshev}.
The precision using the zeroth order DMP eigenvalues is shown with solid curves and with the second order eigenvalues with dashed curves.}
\label{fig:simsq precision}
\end{figure*}

As with $\phi$ in eq.~\ref{eq:phi}, $\psi$ is an excellent approximation for $\wh\theta_{12}$.
The fractional precision of the eigenvalues at zeroth and second order are shown in fig.~\ref{fig:eigenvalue precision}.
Since $\phi$ and $\psi$ are good approximations for $\wh\theta_{13}$ and $\wh\theta_{12}$ respectively, and since $\wh\theta_{23}$ and $\wh\delta$ don't vary very much in matter, one could imagine using the vacuum probabilities with the approximate eigenvalues and replacing only $\theta_{13}$ and $\theta_{12}$ with $\phi$ and $\psi$ respectively.
This is exactly the DMP approach at zeroth order.
Thus one way to quantify the improvement of this approach over DMP is to compare the precision with which we can approximate $\wh\theta_{13}$ and $\wh\theta_{12}$ with either $\phi$ and $\psi$ which result from a two-flavor rotation (see eqs.~\ref{eq:phi} and \ref{eq:psi}) or with eqs.~\ref{eq:Uaisq}, \ref{eq:s13m}, and \ref{eq:s12mc13m}.
We have numerically verified that the full three-flavor approach to calculating the mixing angles improves the precision on the mixing angles in matter (and thus the oscillation probabilities) compared with the two-flavor approach that leads to $\phi$ and $\psi$.

Next, we note that for similar reasons that the first order corrections vanish, $\lambda^{(1)}_i=0$, all the odd corrections vanish within the DMP framework\footnote{In fact, the conditions for the odd corrections to vanish can be generally achieved in an arbitrary three or four dimensional Hamiltonian, but not for general higher dimensional Hamiltonians, see the end of appendix \ref{sec:high order eigenvalues}.}.
That is, $\lambda^{(k)}_i=0$ for all $i\in\{1,2,3\}$ and any $k$ odd, see appendix \ref{sec:high order eigenvalues}.
While it would appear that, given a perturbing Hamiltonian $\propto\eps'$ that the precision would converge as $\eps'$, this shows that, in fact, the precision converges considerably faster at $\eps'^2$.
This result had not been previously identified in the literature.

We now compare the precision of sine of the mixing angles and CP violating phase in matter using the approximate eigenvalues through zeroth order and second order to the exact expressions in fig.~\ref{fig:simsq precision}.
Using the zeroth order eigenvalues to evaluate the angles and the phase is quite precise even at zeroth order, at the $1\%$ level or much better.
Adding in the second order corrections dramatically increases the precision by about four orders of magnitude for neutrinos and six orders of magnitude for anti-neutrinos, consistent with the fact that $\eps'$ is $\sim10^{-2}$ in the limit as $E\to\infty$ and $\sim10^{-3}$ in the limit as $E\to-\infty$.
We also see that we recover the exact answers in vacuum, a trait that many approximation schemes do not share \cite{Parke:2019vbs}.

\begin{figure}[b]
\centering
\includegraphics[width=3.0in]{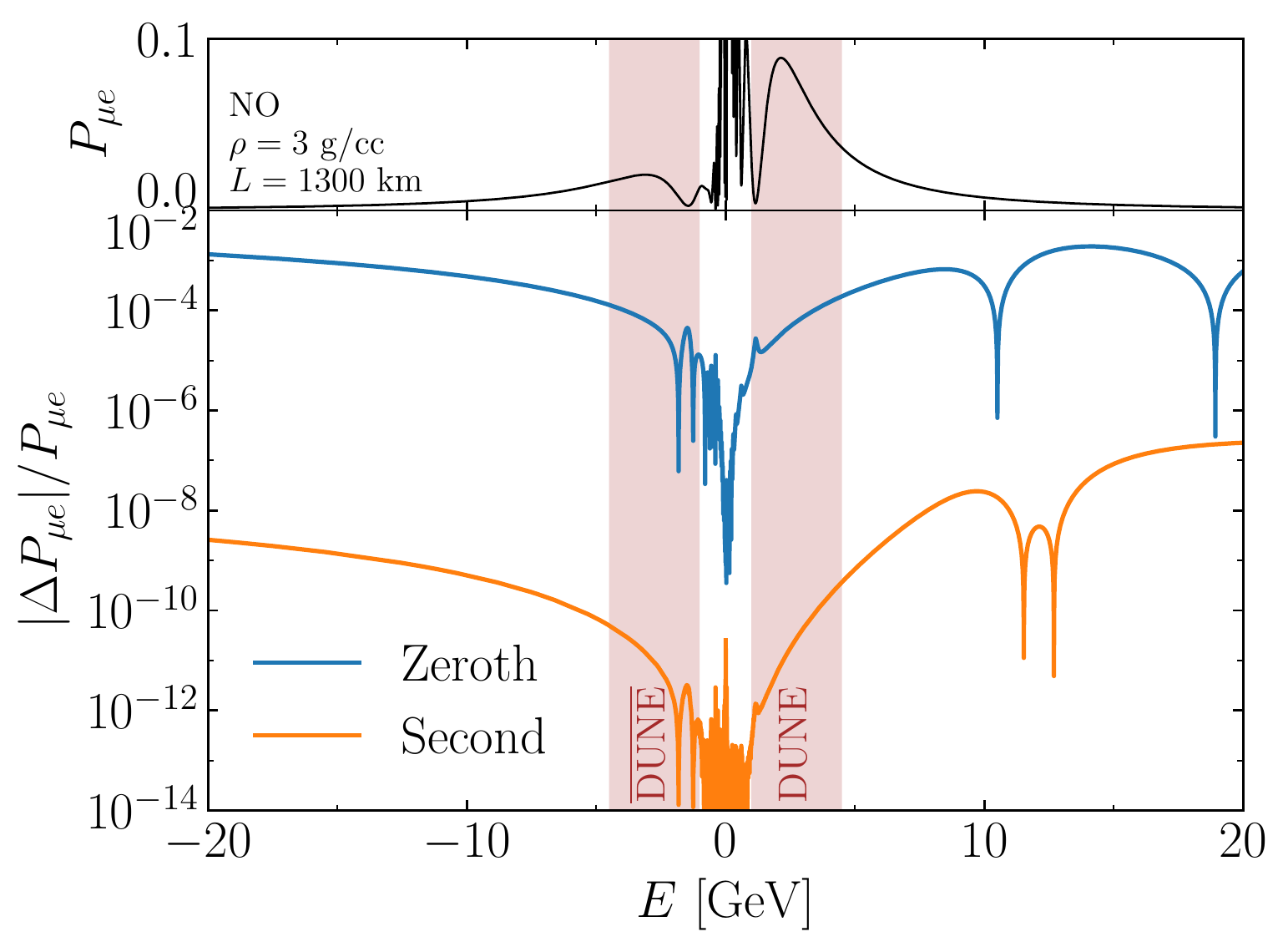}
\caption{{\bf Top}: The oscillation probability $P(\nu_\mu\to\nu_e)$ at $L=1300$ km in the NO.
{\bf Bottom}: The fractional precision of the probability using the zeroth (second) order DMP eigenvalues in blue (orange).
The vertical red bands show DUNE's region of interest.
The precision is comparable in the IO.}
\label{fig:pmue precision}
\end{figure}

Next, we show the precision of the appearance oscillation probability for DUNE in fig.~\ref{fig:pmue precision}\footnote{We also compared calculating the oscillation probabilities with the approximate eigenvalues using the Toshev identity to determine $\delta$ and the NHS identity to determine $\delta$ through the Jarlskog, and found that the Toshev identity performs better.
This is due to the fact that the $\wh\delta$ and $\wh\theta_{23}$ both don't vary very much in matter while all the other parameters do.}.
The scaling law of the precision remains the same as previously shown and we have verified that it continues at the same rate to even higher orders.
In fact, as we continue to higher orders we find that all the odd corrections to the eigenvalues vanish, see appendix \ref{sec:high order eigenvalues}.

\begin{figure}[b]
\centering
\includegraphics[width=3.0in]{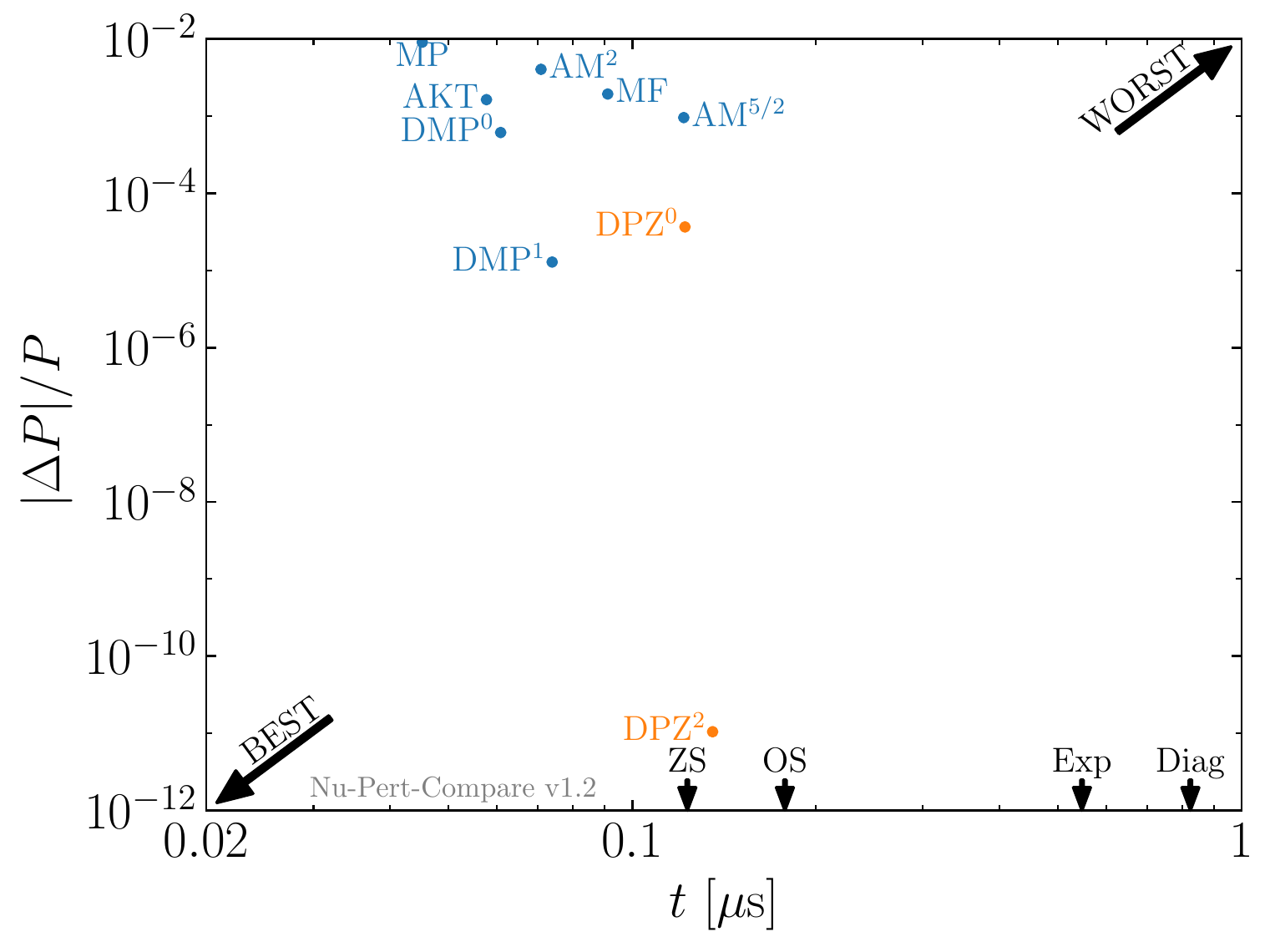}
\caption{We have plotted the fractional precision at the first oscillation maximum for DUNE at $\delta=-0.4\pi$ versus the time to compute one oscillation probability on a single core.
Our results are labeled DPZ and are in orange.
ZS \cite{Zaglauer:1988gz} and Diag are two exact solutions where Diag represents an off-the-shelf linear algebra diagonalization package.
Two other exact solutions from \cite{Ohlsson:1999um} are labeled OS (using the Cayley-Hamilton method) and Exp (exponentiating the Hamiltonian) which do not account for CP violation ($\delta=0$).
We only plot expressions that reach at least 1\% precision at the first oscillation maximum.
The remaining expressions are: MP \cite{Minakata:2015gra}, AM \cite{Asano:2011nj}, MF \cite{Freund:2001pn}, AKT \cite{Agarwalla:2013tza}, and DMP \cite{Denton:2016wmg}.
For a detailed discussion see ref.~\cite{Parke:2019vbs}.}
\label{fig:Speed}
\end{figure}

Finally, in an effort to roughly quantify the ``simplicity'' of our results, we computed the speed with which we can calculate one oscillation probability as shown in fig.~\ref{fig:Speed}.
For comparison we have included many other approximate and exact expressions as previously explored in \cite{Parke:2019vbs}.
For the sake of openness, the \verb.nu-pert-compare. code used for each of these is publicly available \url{https://github.com/PeterDenton/Nu-Pert-Compare} \cite{nu-pert-compare}.
We note that while our new results using DMP eigenvalues are not as fast as others, adding in higher order corrections is extremely simple, as indicated in eqs.~\ref{eq:l12}-\ref{eq:l32} which give rise to an impressive six orders of magnitude improvement in precision for almost no additional complexity.
All points use $\delta=-0.4\pi$ except for OS and Exp where $\delta=0$, for a detailed discussion see ref.~\cite{Parke:2019vbs}.

Computational speed is a useful metric not only for simplicity but also for long-baseline experiments which must compute oscillation probabilities many times when marginalizing over a large number of systematics and standard oscillation parameters.
In addition, performing the Feldman-Cousins method of parameter estimation is known to be extremely computationally expensive \cite{Feldman:1997qc}.

\section{Conclusions}
In this article we have used the eigenvector-eigenvalue identity to develop a new way to write neutrino oscillation probabilities in matter, both exactly and with simpler approximate expressions.
The primary new result involves determining the mixing angles in matter which has the benefit in that intuition gained about vacuum oscillations still applies to oscillations in matter.
The CP violating phase in matter is then determined in a straightforward fashion from $\wh\theta_{23}$ and the Toshev identity \cite{Toshev:1991ku}.
Given the mixing angles and the CP violating phase in matter, writing down the oscillation probability in matter follows directly from the simple vacuum expression.

This technique benefits from the simplicity of using the expression for the oscillation probability in vacuum, and the clear, compact, expressions for the mixing angles in matter, and can be applied to any oscillation channel.
There is also the fact that by explicitly writing the mixing angles as simple functions of the eigenvalues, they can be replaced with simple approximate expressions, such as those derived by Denton, Minakata, and Parke (DMP) \cite{Denton:2016wmg}.
This new technique presented here is more precise than that in DMP, order by order, since this result is effectively complete to all orders in the eigenvectors and only requires correction to the eigenvalues.

The primary new results of this article are,
\begin{itemize}
\item Eq.~\ref{eq:Uaisq}, reproduced here,
\begin{equation}
|\wh U_{\alpha i}|^2=\frac{(\lambda_i-\xi_\alpha)(\lambda_i-\chi_\alpha)}{(\lambda_i-\lambda_j)(\lambda_i-\lambda_k)}\,,\nonumber
\end{equation}
which presents a simple, clear and easy to remember way to determine the norm of the elements of the mixing matrix and hence the mixing angles in matter given the eigenvalues.
Then the oscillation probabilities can be calculated in a straightforward fashion using the CP-violating phase in matter from the Toshev identity.
\item The form of eq.~\ref{eq:Uaisq} allows for the direct substitution of approximate eigenvalues, such as those from DMP.
As shown here for the first time, the DMP eigenvalues converge extremely quickly, $\sim10^{-5}$ per step since all the odd order corrections to the eigenvalues vanish.
\item The form of eq.~\ref{eq:Uaisq} is trivially generalizable to any number of neutrinos.
\end{itemize}
Given the formalism presented here, we have a clear and simple mechanism for calculating the oscillation probabilities in matter either exactly or approximately.
This method hinges on the eigenvalues, therefore the eigenvalues are the Rosetta Stone for neutrino oscillations in matter.

\acknowledgments
We want to thank Terence Tao for many useful and interesting discussions on the eigenvector-eigenvalue identity.
PBD acknowledges the United States Department of Energy under Grant Contract desc0012704 and the Neutrino Physics Center.
This manuscript has been authored by Fermi Research Alliance, LLC under Contract No.~DE-AC02-07CH11359 with the U.S.~Department of Energy, Office of Science, Office of High Energy Physics.
SP received funding/support from the European Unions Horizon 2020 research and innovation programme under the Marie Sklodowska-Curie grant agreement No 690575 and No 674896.

\appendix

\section{The Exact Eigenvalues in Matter}
\label{sec:exact eigenvalues}
From refs.~\cite{cardano,Barger:1980tf,Zaglauer:1988gz}, the exact eigenvalues in matter are,
\begin{align}
\lambda_1&=\frac13 A- \frac13 \sqrt{A^2-3B}\left(S+\sqrt3\sqrt{1-S^2} \right)\,, \nonumber \\
\lambda_2&=\frac13 A- \frac13 \sqrt{A^2-3B}\left(S-\sqrt3\sqrt{1-S^2}\right)\,, \nonumber \\
\lambda_3&=\frac13 A+\frac23\sqrt{A^2-3B}\,S\,.
\end{align}
The terms $A$, $B$, and $C$ are the sum of the eigenvalues, the sum of the products of the eigenvalues, and the triple product of the eigenvalues, while $S$ contains the $\cos(\frac13\cos^{-1}(\cdots))$ terms,
\begin{align}
A&=\Delta m^2_{21}+\Delta m^2_{31}+a\,,\label{eq:A}\\
B&=\Delta m^2_{21}\Delta m^2_{31}+a\left[\Delta m^2_{31}c_{13}^2+\Delta m^2_{21}(1-c_{13}^2s_{12}^2)\right]\,,\label{eq:B}\\
C&=a\Delta m^2_{21}\Delta m^2_{31}c_{13}^2c_{12}^2\,,\label{eq:C}\\
S&=\cos\left\{\frac13\cos^{-1}\left[\frac{2A^3-9AB+27C}{2(A^2-3B)^{3/2}}\right]\right\}\,,\label{eq:S}
\end{align}
where $a\equiv2E\sqrt2G_Fn_e$ is the matter potential, $E$ is the neutrino energy, $G_F$ is Fermi's constant, and $n_e$ is the electron number density.

As an example of the analytic impenetrability of $S$, setting $a=0$ and recovering the vacuum values for the eigenvalues, $(0,\Delta m^2_{21}, \Delta m^2_{31})$, is a highly non-trivial exercise.

\section{Derivation From KTY}
\label{sec:from kty}
Since only the product of elements of the PMNS matrix are necessary to write down the oscillation probabilities, we start with the definition of the product of two elements of the lepton mixing matrix in matter from eq.~39 in KTY, ref.~\cite{Kimura:2002wd},
\begin{equation}
\wh U_{\alpha i}\wh U_{\beta i}^*=\frac{\wh p_{\alpha\beta}\lambda_i+\wh q_{\alpha\beta}-\delta_{\alpha\beta}\lambda_i(\lambda_j+\lambda_k)}{(\lambda_j-\lambda_i)(\lambda_k-\lambda_i)}\,,
\label{eq:UaiUbi}
\end{equation}
where $\wh{x}$ is the quantity $x$ evaluated in matter and the $\lambda_i$'s are the exact eigenvalues in matter\footnote{Note that the eigenvalues in matter can also be expressed in a longer notation as $\lambda_i=\wh{m^2}_i$.}, see appendix \ref{sec:exact eigenvalues}.
We note that a similar approach was used in \cite{Xing:2019owb}.

The matrix $\wh p$ is just the Hamiltonian in matter, $\wh p_{\alpha\beta}=(2E)H_{\alpha\beta}=\sum_i\lambda_i\wh U_{\alpha i}\wh U_{\beta i}^*$, see appendix \ref{sec:hamiltonian}.
The other term, $\wh q$, is given by $\wh q_{\alpha\beta}=\sum_{i<j}\lambda_i\lambda_j\wh U_{\alpha k}\wh U_{\beta k}^*$ for $k\neq i,j$.
It is also equivalent to $\wh q_{\alpha\beta}=(2E)^2(H_{\gamma\beta}H_{\alpha\gamma}-H_{\alpha\beta}H_{\gamma\gamma})$ for $\gamma\neq i,j$.

We evaluate eq.~\ref{eq:UaiUbi} in the case of $\alpha=\beta$.
\begin{equation}
|\wh U_{\alpha i}|^2=\frac{\lambda_i^2-\lambda_i(2E)(H_{\beta\beta}+H_{\gamma\gamma})+\wh q_{\alpha\alpha}}{(\lambda_j-\lambda_i)(\lambda_k-\lambda_i)}\,,
\end{equation}
where $\alpha$, $\beta$, and $\gamma$ are all different, as are $i$, $j$, and $k$.
We can then write the numerator as $(\lambda_i-\xi_\alpha)(\lambda_i-\chi_\alpha)$ where $\xi_\alpha$ and $\chi_\alpha$ satisfy
\begin{align}
\xi_\alpha+\chi_\alpha&=(2E)\left(H_{\beta\beta}+H_{\gamma\gamma}\right)\,, \\
\xi_\alpha\chi_\alpha&=(2E)^2\left(H_{\beta\beta}H_{\gamma\gamma}-H_{\beta\gamma}H_{\gamma\beta}\right)\,.
\end{align}
That is, $\xi_\alpha$ and $\chi_\alpha$ are the eigenvalues of $H_\alpha$, the $2\times2$ submatrix of the Hamiltonian,

We also note that eq.~\ref{eq:Uaisq} leads to the following identity also presented in ref.~\cite{Kimura:2002wd},
\begin{align}
&(\lambda_2-\lambda_1)(\lambda_3-\lambda_1)(\lambda_3-\lambda_2)s_{\wh{12}}c_{\wh{12}}s_{\wh{13}}c_{\wh{13}}^2
\nonumber \\
& =\Delta m^2_{21}\Delta m^2_{31}\Delta m^2_{32}s_{12}c_{12}s_{13}c_{13}^2\,,
\end{align}
which is the NHS identity \cite{Naumov:1991ju,Harrison:1999df} divided by the Toshev identity \cite{Toshev:1991ku}.
That is, this quantity
\begin{equation}
(\lambda_2-\lambda_1)(\lambda_3-\lambda_1)(\lambda_3-\lambda_2)|\wh U_{e1}||\wh U_{e2}||\wh U_{e3}|\,,
\end{equation}
is independent of the matter potential.

\section{The Hamiltonian}
\label{sec:hamiltonian}
Here we multiply out the Hamiltonian in matter for use in the above expressions.
First, we define $(2E)\h=O_{13}(\theta_{13})O_{12}(\theta_{12})M^2O_{12}^\dagger(\theta_{12})O_{13}^\dagger(\theta_{13})+\diag(a,0,0)$, a real matrix, which is,
\begin{widetext}
\begin{equation}
\h=\frac1{2E}
\begin{pmatrix}
a+\Dmsqee s_{13}^2+\Delta m^2_{21}s_{12}^2&\quad&c_{13}s_{12}c_{12}\Delta m^2_{21}&&s_{13}c_{13}\Dmsqee\\[0.5em]
\cdot&&\Delta m^2_{21}c_{12}^2&\quad&-s_{13}s_{12}c_{12}\Delta m^2_{21}\\[0.5em]
\cdot&&\cdot&&\Dmsqee c_{13}^2+\Delta m^2_{21}s_{12}^2
\end{pmatrix}
\,,
\label{eq:Hrot0}
\end{equation}
where $\h_{\alpha\beta}=\h_{\beta\alpha}$ and $\Dmsqee\equiv c_{12}^2\Delta m^2_{31}+s_{12}^2\Delta m^2_{31}$ \cite{Nunokawa:2005nx}.
Then the Hamiltonian in the flavor basis is $H=U_{23}(\theta_{23},\delta)\h U_{23}^\dagger(\theta_{23},\delta)$ which is,
\begin{equation}
H=
\begin{pmatrix}
\h_{ee}&\quad&s_{23}e^{-i\delta}\h_{e\tau}+c_{23}\h_{e\mu}&&c_{23}\h_{e\tau}-s_{23}e^{i\delta}\h_{e\mu}\\[2em]
\cdot&&
\begin{matrix}
c_{23}^2\h_{\mu\mu}+s_{23}^2\h_{\tau\tau}+\\2s_{23}c_{23}\cos\delta\h_{\mu\tau}
\end{matrix}
&\quad&
\begin{matrix}
e^{i\delta}\left[s_{23}c_{23}(\h_{\tau\tau}-\h_{\mu\mu})+\right.\\\left.(c_{23}^2e^{-i\delta}-s_{23}^2e^{i\delta})\h_{\mu\tau}\right]
\end{matrix}\\[2em]
\cdot&&\cdot&&
\begin{matrix}
c_{23}^2\h_{\tau\tau}+s_{23}^2\h_{\mu\mu}-\\2s_{23}c_{23}\cos\delta\h_{\mu\tau}
\end{matrix}
\end{pmatrix}\,,
\end{equation}
where $H_{\alpha\beta}=H_{\beta\alpha}^*$.

Then the eigenvalues of the submatrices, $\xi_\alpha$ and $\chi_\alpha$ are given by
\begin{align}
\xi_e+\chi_e&=(2E)\left(\h_{\mu\mu}+\h_{\tau\tau}\right)\,,\label{eq:sub e sum}\\
\xi_e\chi_e&=(2E)^2\left(\h_{\mu\mu}\h_{\tau\tau}-\h_{\mu\tau}^2\right)\,,\label{eq:sub e prod}\\
\xi_\mu+\chi_\mu&=(2E)\left(\h_{ee}+c_{23}^2\h_{\tau\tau}+s_{23}^2\h_{\mu\mu}-2s_{23}c_{23}\cos\delta\h_{\mu\tau}\right)\,,\label{eq:sub mu sum}\\
\xi_\mu\chi_\mu&=(2E)^2\left[\h_{ee}\left(c_{23}^2\h_{\tau\tau}+s_{23}^2\h_{\mu\mu}-2s_{23}c_{23}\cos\delta\h_{\mu\tau}\right)-\left|c_{23}\h_{e\tau}-s_{23}e^{-i\delta}\h_{e\mu}\right|^2\right]\,.\label{eq:sub mu prod}
\end{align}
The $\xi_\tau$ and $\chi_\tau$ eigenvalues are the same as $\xi_\mu$ and $\chi_\mu$ under the interchange $s_{23}^2\leftrightarrow c_{23}^2$ and $s_{23}c_{23}\to-s_{23}c_{23}$.
Note that the complicated $H_{\mu\tau}$ term does not appear in the $\xi_e$ and $\chi_e$ terms since the eigenvalues of the $2\times2$ submatrix $H_e$ are the same as those of $\h_e$.
\end{widetext}

For illustration, we write down the electron submatrix eigenvalues, although we note that explicit calculation of the submatrix eigenvalue is not necessary since eq.~\ref{eq:Uaisq} depends only on the sum and product of the eigenvalues which are directly given in eqs.~\ref{eq:sub e sum}-\ref{eq:sub mu prod},
\begin{align}
\xi_e,\chi_e&=\frac{\Dmsqee}2\left[c_{13}^2+\eps\pm\sqrt{(c_{13}^2-\eps)^2+(2s_{13}s_{12}c_{12}\eps)^2}\right]\,, \nonumber \\
\end{align}
where $\eps\equiv\Delta m^2_{21}/\Dmsqee$.
(This $\epsilon$ is different from\\ $\epsilon^\prime = \sin(\phi-\theta_{13}) s_{12} c_{12} \Delta m^2_{21}/\Dmsqee$ used as our perturbative expansion parameter, see eq.~\ref{eq:epsp}.)

\section{Asymptotics of Submatrix Eigenvalues}
\label{sec:submatrix asymptotics}
For convenience we define $\xi_\alpha<\chi_\alpha$.
Then we note that the submatrix eigenvalues are asymptotically the same as certain full eigenvalues.
\begin{align}
\lim_{E\to-\infty}\lambda_2&=\lim_{E\to\infty}\lambda_1=\xi_e\,,\\
\lim_{E\to-\infty}\lambda_3&=\lim_{E\to\infty}\lambda_2=\chi_e\,,\\
\lim_{E\to-\infty}\lambda_1&=\lim_{E\to-\infty}\xi_\mu=\lim_{E\to-\infty}\xi_\tau\,,\\
\lim_{E\to\infty}\lambda_3&=\lim_{E\to\infty}\chi_\mu=\lim_{E\to\infty}\chi_\tau\,,\\
\lim_{E\to-\infty}\chi_\mu&=\lim_{E\to\infty}\xi_\mu=m_\tau^2\,,\\
\lim_{E\to-\infty}\chi_\tau&=\lim_{E\to\infty}\xi_\tau=m_\mu^2\,,
\end{align}
where
\begin{equation}
m_\alpha^2=\sum_im_i^2|U_{\alpha i}|^2\,.
\end{equation}
Since $m_\mu^2\approx c_{13}^2s_{23}^2\Delta m^2_{31}$ and $m_\tau^2\approx c_{13}^2c_{23}^2\Delta m^2_{31}$, changing the octant changes which of $m_\mu^2$ and $m_\tau^2$ are larger.
This in turn swaps the ordering of the $\mu$ and $\tau$ submatrix eigenvalues.

Furthermore, the eigenvalues of $H$ and its principal minors, $\lambda_{i}$'s and $\xi_\alpha$'s, $\chi_\alpha$'s ($\xi_\alpha<\chi_\alpha$), satisfy the Cauchy interlacing identity, 
$\lambda_1 \leq \xi_\alpha \leq \lambda_2 \leq \chi_\alpha \leq\lambda_3$ for $\alpha =(e,\mu,\tau)$, for all values of the matter potential.

\newcommand{\dms}{\Delta m_{21}^2}
\newcommand{\dma}{\Delta m_{31}^2}
\newcommand{\dmb}{\Delta m_{32}^2}
\newcommand{\sz}{s_{12}}
\newcommand{\sn}{s_{23}}
\newcommand{\st}{s_{13}}
\newcommand{\cz}{c_{12}}
\newcommand{\cn}{c_{23}}
\newcommand{\ct}{c_{13}}

\begin{table*}[t]
\center
\begin{tabular}{|c|c|c|}
\hline   &  $\xi^\prime+ \chi^\prime$  & $\xi^\prime \chi^\prime$  \\
\hline  e &   $\dma\ct^2 + \dms(1-\ct^2 \sz^2) $ & $(\dma \ct^2)(\dms \cz^2)$\\
 $\mu$  & $ a+ \dma  +  \dms \sz^2 $   & $a (\dma \ct^2+\dms \sz^2 \st^2)+\dma \dms \sz^2 $ \\
  $\tau$ \quad  & \quad  $ a+\dma \st^2 + \dms (1-\st^2 \sz^2)$    \quad  & \quad $a \dms \cz^2 +( \dma \st^2)( \dms \cz^2 ) $ \quad \\
  \hline Sum &  $2(a+ \dma  +  \dms)=2A $   &~~  $a (\dma \ct^2+\dms (1- \ct^2 \sz^2))+\dma \dms=B  $ ~~ \\
  \hline
  \end{tabular}
  \caption{The sum and product of the eigenvalues of the  principal minors of the rotated Hamiltonian, eq.~\ref{eq:Hrot} and their relationship to eigenvalues of the full Hamiltonian, $\sum_j \lambda_j(H)=A$ and $\sum_{j>k} \lambda_j(H) \lambda_k(H)=B$ given in appendix \ref{sec:exact eigenvalues}.
  }
  \label{tab:xichiprime}
  \end{table*}

\begin{widetext}

\section{Using the $(\theta_{23}, \delta)$-Rotated Flavor Basis}
\label{sec:rotated basis}
In this appendix, we use the eigenvector-eigenvalue identity in the vacuum  $(\theta_{23}, \delta)$-rotated flavor basis and recover the full PMNS matrix in matter by performing the vacuum  $(\theta_{23}, \delta)$-rotated at the end. The $(\theta_{23}, \delta)$-rotated flavor basis is defined as
\begin{align}
U^\dagger_{23} (\theta_{23}, \delta) 
\begin{pmatrix} \nu_e \\ \nu_\mu \\ \nu_\tau \end{pmatrix}  
\quad \text{with}
\quad
&U_{23}(\theta_{23}, \delta)  = \begin{pmatrix}
1\\&c_{23}&s_{23}e^{i\delta}\\&-s_{23}e^{-i\delta}&c_{23}
\end{pmatrix}.
\end{align}
In this basis the Hamiltonian, $\h$, is given as
\begin{equation}
\h=\frac1{2E}
\begin{pmatrix}
a+\Dmsqee s_{13}^2+\Delta m^2_{21}s_{12}^2&\quad&c_{13}s_{12}c_{12}\Delta m^2_{21}&&s_{13}c_{13}\Dmsqee\\[0.5em]
\cdot&&\Delta m^2_{21}c_{12}^2&\quad&-s_{13}s_{12}c_{12}\Delta m^2_{21}\\[0.5em]
\cdot&&\cdot&&\Dmsqee c_{13}^2+\Delta m^2_{21}s_{12}^2
\end{pmatrix} \,.
\label{eq:Hrot}
\end{equation}
Note, it is now real and independent of $\theta_{23}$ and $\delta$ and the same as eq.~\ref{eq:Hrot0}.
\end{widetext}

This Hamiltonian can be diagonalized by the following unitary matrix
\begin{align}
\wh{V} \equiv \wh{V}_{23}({\alpha})~\wh{U}_{13}~ \wh{U}_{12} \, ,
\end{align}
with only real entries, with\footnote{The relationship between  this work and ref.~\cite{Zaglauer:1988gz} is  $\sin^2 {\alpha} = F^2/(E^2+F^2)$ and $F^2=(\lambda_3-\xi^\prime_e)(\lambda_3-\chi^\prime_e)(\lambda_3-\xi^\prime_\mu)(\lambda_3-\chi^\prime_\mu)$ and $E^2=(\lambda_3-\xi^\prime_e)(\lambda_3-\chi^\prime_e)(\lambda_3-\xi^\prime_\tau)(\lambda_3-\chi^\prime_\tau)$. The characteristic equation, using $\lambda_3$ as the solution, 
 is needed to prove this equivalence.}
\begin{align}
s_{\wh{13}}^2=|\wh V_{e 3}|^2&=\frac{(\lambda_3-\xi^\prime_e)(\lambda_3-\chi^\prime_e)}{(\lambda_3-\lambda_1)(\lambda_3-\lambda_2)}\,, \notag \\
s_{\wh{12}}^2c_{\wh{13}}^2=|\wh V_{e 2}|^2&=-\frac{(\lambda_2-\xi^\prime_e)(\lambda_2-\chi^\prime_e)}{(\lambda_2-\lambda_1)(\lambda_3-\lambda_2)}\,, \notag \\
s_{{\alpha}}^2c_{\wh{13}}^2=|\wh V_{\mu 3}|^2&=\frac{(\lambda_3-\xi^\prime_\mu)(\lambda_3-\chi^\prime_\mu)}{(\lambda_3-\lambda_1)(\lambda_3-\lambda_2)}\,,\label{eq:salpha}
\end{align}
where the sum and the product of $\xi^\prime$'s and $\chi^\prime$'s are given in Table \ref{tab:xichiprime} and are obtained from the trace and determinant of the  principal minors of $\h$, eq.~\ref{eq:Hrot}.  The expressions for $\wh{\theta}_{12}$ and $ \wh{\theta}_{13}$ are the same as eqs.~\ref{eq:s12mc13m} and \ref{eq:s13m}, since $\xi^\prime_e=\xi_e$ and $\chi^\prime_e=\chi_e$, eqs.~\ref{eq:sub e sum} and \ref{eq:sub e prod}.  
$|\alpha|$ is tiny ($< 0.01$) and is zero in vacuum.

For the full PMNS matrix in matter, we combine $U_{23}(\theta_{23}, \delta)$ with $\wh{V}_{23}({\alpha})$ into $U_{23}(\wh{\theta}_{23}, \wh{\delta})$, as follows
\begin{align}
& \left( \begin{array}{cc}
c_{23}  & s_{23} e^{i\delta} \\
 -s_{23} e^{-i\delta }& c_{23} 
 \end{array}
 \right)
 ~\left( \begin{array}{cc}
c_{\alpha}  & s_{\alpha}  \\
 -s_{\alpha}  & c_{\alpha} 
 \end{array}
 \right)
\notag  \\[3mm]
& =  \left( \begin{array}{cc}
e^{i\rho} &  \\
 &  e^{i\sigma}
 \end{array}
 \right)
 \left( \begin{array}{cc}
c_{\wh{23}}  & s_{\wh{23}} e^{i\wh{\delta}} \\
 -s_{\wh{23}} e^{-i\wh{\delta} }& c_{\wh{23}} 
 \end{array}
\right) \,.
 \end{align}
 The solution is $\sigma =-\rho$ with
 \begin{align}
 c_{23} c_\alpha -s_{23}s_\alpha  e^{i\delta} &= c_{\wh{23}} e^{i\rho}\, , \notag  \\
 \text{and}  \quad s_\alpha c_{23} +c_\alpha s_{23} e^{i\delta} & =  s_{\wh{23}} e^{i(\rho+\wh{\delta})}  \notag \, .
 \end{align}
Therefore
\begin{align}
s^2_{\wh{23}} &= c^2_\alpha s^2_{23} +s^2_\alpha c^2_{23} + 2 s_\alpha c_\alpha s_{23} c_{23} \cos \delta \,,  \label{eq:s23m}  \\
\cos \wh{\delta}& =(\cos \delta \, s_{23}c_{23}(c^2_\alpha-s^2_\alpha)  +(c^2_{23}-s^2_{23})s_\alpha c_\alpha )/ ( s_{\wh{23}} c_{\wh{23}} ) \notag  \, ,\\
\sin \wh{\delta} & = (\sin \delta \, s_{23}c_{23} )/ ( s_{\wh{23}} c_{\wh{23}} )\,.
\label{eq:deltam} 
\end{align}
The equation for $\sin \wh{\delta}$ is exactly the Toshev identity \cite{Toshev:1991ku}. 
 
  The full PMNS matrix in matter is then by
 \begin{align}
\wh{U}_{PMNS}  = U_{23}(\wh{\theta}_{23}, \wh{\delta})~U_{13}(\wh{\theta}_{13} )~ U_{12}(\wh{\theta}_{12})\, .
\end{align}
with $\wh{\theta}_{23}$, $\wh{\delta}$, $\wh{\theta}_{12}$ and $ \wh{\theta}_{13}$ given by eqs.~\ref{eq:s23m}, \ref{eq:deltam},  \ref{eq:s12mc13m}, and \ref{eq:s13m}, respectively, in total agreement with what was obtained in ref.~\cite{Zaglauer:1988gz}, using a different method.

\section{Extension to an Arbitrary Number of Neutrinos}
\label{sec:extension arbitrary}
Equation \ref{eq:Uaisq} can be generalized in a straightforward fashion to an arbitrary number of neutrinos.
As an initial illustrative example, for two flavors in matter we have that the elements of the diagonalized mixing matrix in matter are
\begin{equation}
|\wh U_{\alpha i}|^2=\frac{\lambda_i-\xi_\alpha}{\lambda_i-\lambda_j}\,.
\end{equation}
(This two-flavor approach was exploited in a three-flavor context in DMP via two two-flavor rotations.)

To evaluate this, we find the eigenvalues of the Hamiltonian and its submatrix.
The Hamiltonian is
\begin{equation}
H=\frac{\Delta m^2}{4E}
\begin{pmatrix}
a/\Delta m^2-\cos2\theta&\sin2\theta\\
\sin2\theta&\cos2\theta-a/\Delta m^2
\end{pmatrix}\,.
\end{equation}
The eigenvalues, $\lambda_{1,2}/2E$, of this $2\times2$ system are,
\begin{equation}
\lambda_{1,2}=\mp\frac12\sqrt{(a-\Delta m^2\cos2\theta)^2+(\Delta m^2\sin2\theta)^2}\,,
\end{equation}
and the submatrix eigenvalues, $\xi_{e,\mu}/2E$ are trivially,
\begin{equation}
\xi_e=\frac12\left(\Delta m^2\cos2\theta-a\right)\,,\quad
\xi_\mu=\frac12\left(a-\Delta m^2\cos2\theta\right)\,.
\end{equation}
Then we can write down the mixing matrix in matter, where we note that the off-diagonal term squared is $\sin^2\wh\theta$,
\begin{align}
\sin^2\wh\theta&=|\wh U_{e2}|^2=\frac{\lambda_2-\xi_e}{\lambda_2-\lambda_1} \\
&=\frac12\left(1-\frac{\Delta m^2\cos2\theta-a}{\sqrt{(\Delta m^2\cos2\theta-a)^2+(\Delta m^2\sin2\theta)^2}}\right)\,,  \nonumber 
\end{align}
which agrees with the standard two-flavor expression \cite{Parke:1986jy}.

What is powerful about this method is that it can extend to more neutrinos as well.
For four neutrinos, we can write down the elements of the mixing matrix in matter in the related, easy to remember form as the two or three-flavor cases,
\begin{equation}
|\wh U_{\alpha i}|^2=\frac{(\lambda_i-\xi_\alpha)(\lambda_i-\chi_\alpha)(\lambda_i-\zeta_\alpha)}{(\lambda_i-\lambda_j)(\lambda_i-\lambda_k)(\lambda_i-\lambda_\ell)}\,,
\end{equation}
for $i,j,k,\ell$ all different and where $\xi_\alpha$, $\chi_\alpha$, and $\zeta_\alpha$ are the three eigenvalues of the associated submatrix.
This method can be extended in a straightforward fashion to an arbitrary number of neutrinos.

The general form of eq.~\ref{eq:Uaisq} for any $n\times n$ matrix with possibly degenerate eigenvalues is,
\begin{equation}
|\wh U_{\alpha i}|^2\prod_{k=1,k\neq i}^n(\lambda_i-\lambda_k)=\prod_{k=1}^{n-1}(\lambda_i-\xi_{\alpha,k})\,,
\end{equation}
where the $k$ in $\xi_{\alpha,k}$ covers the $n-1$ eigenvalues of the $\alpha$ submatrix, see \cite{Denton:2019pka}.

\section{Higher Order Eigenvalues}
\label{sec:high order eigenvalues}
As mentioned in the text, the first order corrections to the eigenvalues are zero.
The second order corrections are given in eqs.~\ref{eq:l12}-\ref{eq:l32}.
We also find that the third order corrections are zero.

Given an $3\times 3$ Hamiltonian
\begin{equation}
H=H_0+V\,,
\end{equation}
where $H_0=\text{diag}(\lambda_1,\,\lambda_2,\,\lambda_3)$ is a non-degenerate zeroth order diagonal matrix and $V$, which is Hermitian, is the perturbing part of the Hamiltonian in DMP,
\begin{equation}
V=\eps'\frac{\Delta m^2_{ee}}{2E}
\begin{pmatrix}
& &-s_{\psi} \\
& &c_\psi \\
-s_\psi&c_\psi &
\end{pmatrix}\,.
\label{eq:hcheck1}
\end{equation}
Because all diagonal elements of $V$ vanish, it is straightforward to see that the first order eigenvalue corrections are zero. Next we calculate the third order corrections to eigenvalues
\begin{equation}
\lambda_i^{(3)}=\sum_{j,k\neq i}\frac{V_{ij}V_{jk}V_{ki}}{(\lambda_j-\lambda_i)(\lambda_k-\lambda_i)}-\sum_{j\neq i}\frac{V_{ii}V_{ij}V_{ji}}{(\lambda_j-\lambda_i)^2}\,,
\end{equation}
where $V_{ij}=\langle i|V|j\rangle$.
The rightmost term is clearly zero since $V_{ii}=0$.
In the first summation if $j=k$, $V_{jk}=0$ for the same reason.
If not then the numerator contains the product of all three off-diagonal terms.
Since one of these terms is zero in DMP ($V_{12}$) we have that $\lambda_i^{(3)}$ is zero.

In fact, all $\lambda_i^{(m)}=0$ for $m$ odd.
A brief proof is that when $m$ is odd $\lambda^{(m)}_i$ is a summation of terms proportional either to a diagonal element of $V$ or the product $V_{12}V_{23}V_{31}$. 

The above conclusion is a special case of a more general statement.
Let's consider an $n\times n$ perturbing Hamiltonian $V_{n\times n}$ for which there is an $r$ such that $0\le r\le n$ wherein
\begin{equation}
V_{ij}\neq0\quad\text{only if}\quad i\leq r<j\quad\text{or}\quad i>r\geq j\,,
\end{equation}
i.e.~$V_{n\times n}$ has the block form
\begin{equation}
V_{n\times n}=
\begin{pmatrix}
0_{r\times r} & V_{r\times(n-r)}\\
V_{(n-r)\times r} &0_{(n-r)\times(n-r)}
\end{pmatrix}\,.
\end{equation}
If the above condition is satisfied, the $m$th order eigenvalue corrections $\lambda^{(m)}_i=0$ for $m$ odd.
DMP is the case of $n=3$, $r=2$.

We list the even order corrections through 6$^{\rm th}$ order.
We note that we only need to write down the $\lambda_1^{(n)}$ corrections since $\lambda_2^{(n)}$ is related to $\lambda_1^{(n)}$ by the 1-2 interchange symmetry \cite{Denton:2016wmg} as given by eq.~\ref{eq:12 interchange symmetry} and $\sum_i\lambda_i^{(n)}=0$ which allows for the determination of $\lambda_3^{(n)}$.
\begin{widetext}
\begin{align}
\frac{\lambda^{(2)}_1}{(\eps'\Dmsqee)^2}={}&-\frac{s_\psi^2}{\Delta\lambda_{31}}\,,\\
\frac{\lambda^{(4)}_1}{(\eps'\Dmsqee)^4}={}&\frac{s_\psi^2}{\Delta\lambda_{21}(\Delta\lambda_{31})^3}\left(-c_\psi^2\Delta\lambda_{31}+s_\psi^2\Delta\lambda_{21}\right)\,,\\
\frac{\lambda^{(6)}_1}{(\eps'\Dmsqee)^6}={}&\frac{s_\psi^2}{(\Delta\lambda_{21})^2(\Delta\lambda_{31})^5}\left[-c_\psi^4(\Delta\lambda_{31})^2+c_\psi^2s_\psi^2(3\Delta\lambda_{21}+\Delta\lambda_{31})\Delta\lambda_{31}-2s_\psi^4(\Delta\lambda_{21})^2\right]\,.
\end{align}
\end{widetext}
For three neutrinos, the Jacobi method ensures that the conditions that the odd corrections to the eigenvalues can always be met by rotating one off-diagonal element of the perturbing Hamiltonian to zero.
The necessary conditions can also be met for four neutrinos by rotating two off-diagonal elements in disconnected sectors (say, $U_{12}$ and $U_{34}$).
For general matrices this condition cannot be met for more than four neutrinos.

\bibliography{Rosetta}

\end{document}